\documentclass[11pt]{article}
\usepackage{moriond,epsfig}

\def\be{\begin{equation}}
\def\ee{\end{equation}}
\def\bea{\begin{eqnarray}}
\def\eea{\end{eqnarray}}

\begin{document}
\begin{flushright}
CERN-PH-TH/2006-123\\
Fermilab-CONF-06-222-T
\end{flushright}
\vspace*{4cm}
\title{CLOSING TALK: QCD MORIOND 2006}

\author{R. KEITH ELLIS}

\address{Theory Department, Fermilab, PO Box 500, Batavia, IL 60510, USA\\
and\\
TH Department, CERN, 1211 Geneva 23, Switzerland }

\maketitle\abstracts{
I comment on some of the theoretical work presented at QCD Moriond, 2006}

\def\cpaper#1{\vspace*{-0.02cm}\hfill\magenta{{\tiny \hbox{#1}}}\vspace*{0.02cm}}
\def\cpaperr#1{\vspace*{-0.02cm}\hfill\red{{\tiny \hbox{#1}}}\vspace*{0.02cm}}
\def\cpaperb#1{\vspace*{-0.02cm}\hfill\blue{{\tiny \hbox{#1}}}\vspace*{0.02cm}}
\def\cpaperleft#1{\vspace*{-0.02cm}\magenta{{\tiny \hbox{#1}}}\hfill\vspace*{0.02cm}}
\def\bit{\begin{itemize}}
\def\eit{\end{itemize}}
\def\as{\alpha_{S}}
\def\beqn{\begin{eqnarray}}
\def\eeqn{\end{eqnarray}}
\def\beq{\begin{equation}}
\def\eeq{\end{equation}}
\def\slsh{\rlap{$\;\!\!\not$}}
\newcommand{\vev}[1]{\langle{#1}\rangle}
\def\VEV#1{\left\langle #1\right\rangle}

One of the advantages of giving the closing talk at a conference with only 
plenary sessions, is that a summary is certainly superfluous. I shall take 
full advantage of that freedom in my talk and present only a few topics.
\section{QCD at High Energy}
%\begin{figure}
%\begin{minipage}{7cm}
%\includegraphics[scale=0.3]{glass_saturation.eps}
%\caption{IMF view of proton}
%\end{minipage}
%\hfill
%\begin{minipage}{7cm}
\begin{figure}
\begin{center}
\includegraphics[scale=0.5]{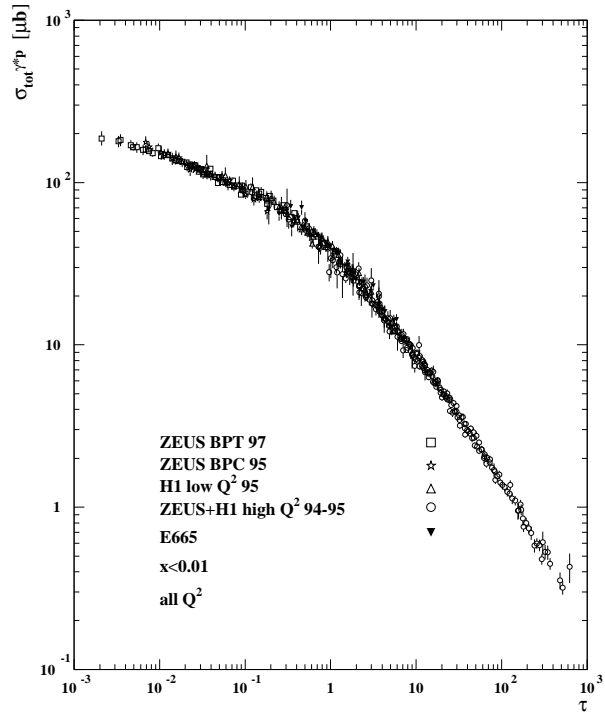}
\caption{Total deep inelastic scattering cross section vs. $\tau$ [1].}
\label{gs}
%\end{minipage}
\end{center}
\end{figure}
In the infinite momentum frame (IMF) a proton or nucleus is seen as a 
Lorentz contracted disk, and $x g(x,Q^2)$ is the number  of gluons per unit rapidity
of transverse size less than $1/Q$ on that disk. 
At small $x$ the number of gluons becomes so large that 
there is a saturation in 
the growth of the number of gluons. 
As a consequence of this 
saturation, the total virtual-photon proton scattering cross 
section satisfies a geometrical scaling law~\cite{Stasto:2000er} as shown in Fig.~\ref{gs},
\beq
\sigma(x,Q^2) \to \sigma(\tau), \;\; \tau=Q^2/Q_s^2(x),\;\; 
Q_s(x)=(x/x_0)^\lambda,\;\;\lambda=0.3\;\;.
\eeq
$Q_s(x)$ is the $x$-dependent saturation scale.
\subsection{Recent developments}
\begin{figure}
\begin{center}
\includegraphics[scale=0.45]{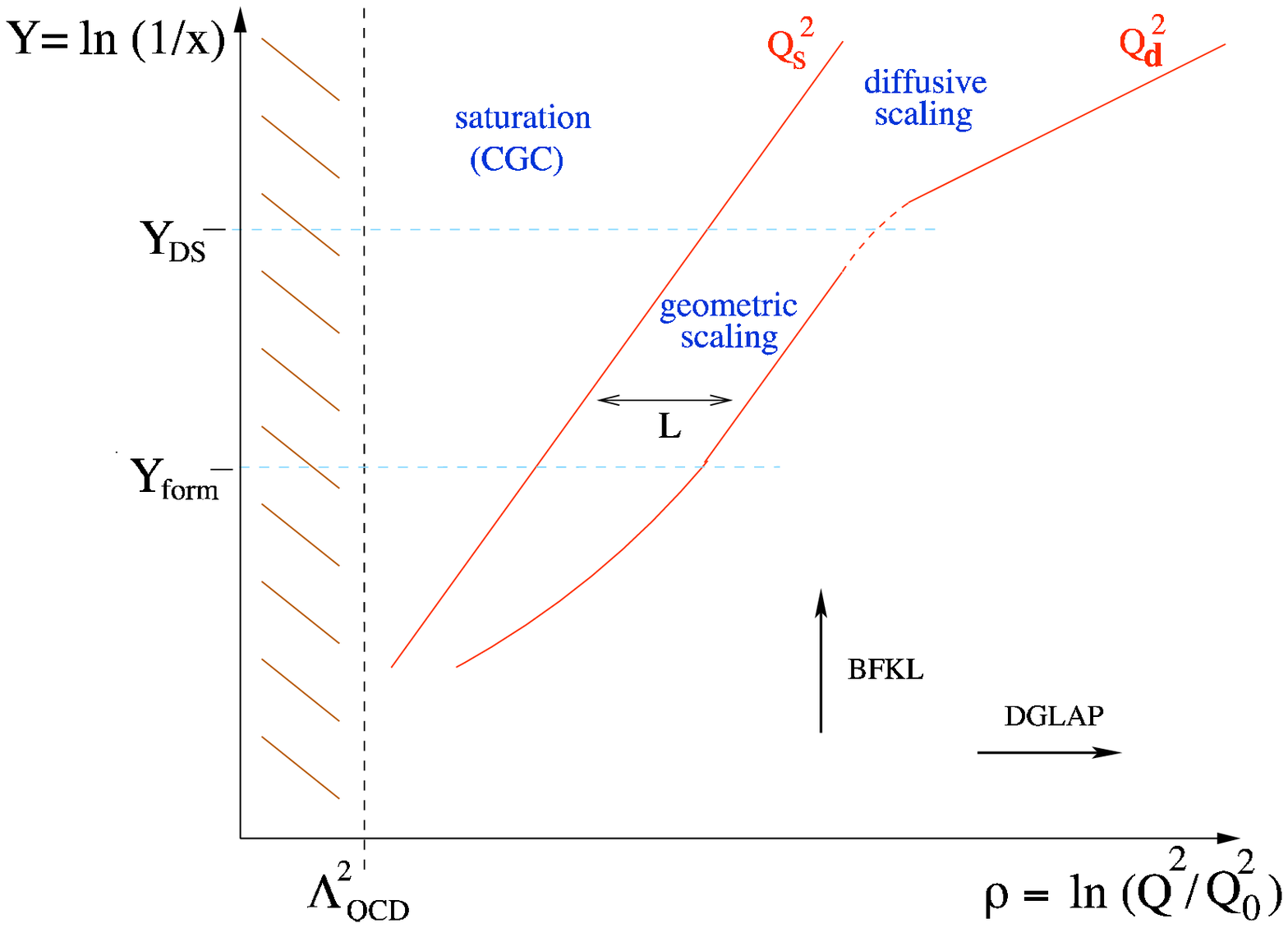}
\caption{A modern view of the phase diagram of QCD [2].}
\label{iancu}
\end{center}
\end{figure}
Recent work in this area is beginning to look in more detail at the 
region near and beyond the saturation boundary.~\cite{Iancu,Lublinsky:2006cb}
While there is geometrical scaling in the region close to the saturation 
boundary, there are expectations that as one proceeds to small $x$ along the
saturation boundary one should arrive at a region of diffusive scaling
\beq
\sigma(x,Q^2) \to \sigma\Big(\frac{\ln \tau}{\sqrt{Y}}\Big), \;\; Y=\ln(1/x).
\eeq 
Fig.~\ref{iancu} gives a modern view of the phase diagram of QCD at small $x$. 
The interest in the region beyond the saturation boundary,
(the region of the colour glass condensate), 
stems from the fact that it is a region of large occupation number, 
but of weak 
coupling. Hence it is amenable to perturbative treatment.
% Numerical investigations with a transverse lattice
%of CGC region.~\cite{Gelis}

\section{DIS scattering in the rest frame}
The description of deep inelastic scattering (DIS) in the rest frame of the target 
hadron has been understood for many years.~\cite{Gribov,bj}
In this frame the average lifetime of the fluctuation of a virtual photon
into a $q \bar{q}$-pair is $\Delta t \sim 1/\Delta E \sim 1/(M x_B)$.
Thus at small Bjorken $x_B$, the virtual photon materializes into a 
$q \bar{q}$ pair upstream of the target
and arrives at the face of nucleon or nucleus in hadronic guise.
The dipoles which interact are either small with symmetric longitudinal momenta
or large with asymmetric longitudinal momenta, so that the cross-section
falls like $1/Q^2$ as required by Bjorken scaling.

\subsection{Dipole scattering formula}
Thus in the target rest frame the high-energy cross section is 
the product of the wave function of the virtual photon
to produce a dipole times the cross section for the dipole to interact.
The cross section for the scattering of a longitudinal or 
transverse photon is
\beq
\sigma_{L/T}(x,Q^2) = \int dz \; d^2 {\bf r} |\Psi_{L/T}(z,r)|^2 \hat{\sigma}(x,r)\;,
\eeq
where the dipole cross section is related to the unintegrated gluon distribution function
of the IMF description, ${\cal F}(x,{\bf k})$,
\beq
\hat{\sigma}(x,r)=\frac{4\pi \as}{3} \int \frac{d^2 {\bf k}}{{\bf k}^4} \; 
{\cal F}(x,{\bf k})
\big(1-\exp (i {\bf k} \cdot {\bf r})\big)\; .
\eeq

The  phenemenological Golec-Biernat-Wusthoff model~\cite{Golec-Biernat:1998js} 
attempts to describe the dipole cross section for all $r$.
At small $r$, the cross section $\sigma \sim r^2$, which is the mathematical expression of the
phenomenon of colour transparency.
At large $r$, 
the dipole cross section is bounded by an energy independent
value $\sigma_0$ which assures the unitarity of $F_2$,
\beq
\hat{\sigma}(r,x)\, =\, \sigma_0\,
\left\{1-\exp\left(-\frac{Q_s^2 (x)r^2}{4}\right)\right\}\;.
\eeq
The transition between the two regimes is controlled by the 
saturation scale $Q_s(x)$, which increases with increasing $1/x$. 
The GBW model can be improved by several refinements,
such as the inclusion of impact parameter dependence
and the incorporation of perturbative evolution in the dipole cross section.~\cite{Kowalski} 
In addition, heavy flavours can be included in the wave function of the 
virtual photon.~\cite{Sapeta:2006th} 

The advantage of this approach is that 
the model of a dipole interacting with a nucleus at rest
also gives information about the structure of 
jets and rapidity gaps in the event. 
This in turn can be used to make predictions for diffraction, shadowing,
and vector meson production.
Thus we see that the formulation of DIS in the rest frame
gives an appealing physical picture, which has predictivity 
beyond inclusive scattering.
Although, in the end, all physics must be Lorentz invariant, 
the description is less clear in the infinite momentum frame.

\subsection{Diffraction in hadron-hadron scattering}
The models for diffraction in hadron-hadron scattering have not yet been refined 
and informed by data to the extent achieved in DIS.
In an interesting report the CDF collaboration has reported 3 diphoton candidates
in diffractive events.~\cite{Terashi:2006bc} Although the background to this data
has not yet been estimated, such measurements certainly have the potential  
to help in the estimation 
of diffractive production of other particles, such as the Higgs boson.
The issue of the diffractive Higgs cross section 
and hard scattering corrections
certainly needs to be resolved.~\cite{Khoze}
Current estimates of the cross sections for the production of standard 
model Higgs bosons tend to be rather low. 
As described by De Roeck and Khoze~\cite{deRoeck,Khoze} the exclusive diffractive Higgs production 
cross section, $pp \to pHp$ is estimated to 
be 3-10 fb,whereas the inclusive Higgs production cross section, $pp \to pXHYp$ is 50-200fb.
Thus for a standard model Higgs one expects 
11(10) signal(background) events for Higgs production in 30 fb$^{-1}$. 

\section{Spin and $\Delta g$}
The total spin of the proton is made up the contributions of quarks, gluons
and their angular momentum, $\frac{1}{2} = \frac{1}{2} \eta_\Sigma +\eta_g +<L_z>$,
where $\eta_g = \int dx (g^\uparrow-g^\downarrow )$ etc. 
As described by Yuan~\cite{Yuan},
the historical results, $\eta_\Sigma = \eta_u + \eta_d +\eta_s =0.3 \pm 0.1$ 
and $\eta_s = -0.1 \pm 0.03$ are at variance with naive quark model ideas, and have 
stimulated a number of experiments to measure the spin carried by the gluons directly.
The possibility that the spin carried by the gluon might be large is motivated by
the evolution equations for the first moments of the spin-dependent structure functions,
\beq
\frac{d }{d \ln Q^2} \; \eta_\Sigma (t) =0 +O(\alpha_s^2),  \; \;
\frac{d }{d \ln Q^2} \; \as(t) \eta_g (t) = 0 +O(\alpha_s^2) \; ,
\eeq
which indicate that $\alpha_s \times $gluon contribution
is formally of the same order as the quark  contribution.
Early experiments will not be able to make very precise statements about $\eta_g$,
but they should be able to indicate whether a sizeable fraction of the total spin of 
one half is carried by the gluons, and {\it a fortiori}, 
whether the spin carried by the gluons
is so large,
\beq
\eta_g(4~{\rm GeV}^2)=2,\;\; (\eta_s \sim - \frac{\as}{2 \pi } \eta_g)\;, 
\eeq
that it can explain the apparent discrepancy with naive quark model ideas.

At this meeting there were new 
experimental results~\cite{Fatemi:2006aa,Procureur:2006sg,Eyser:2006sm}
sensitive to $\eta_g$,
but the overall picture is unchanged.
The experimental results give little comfort to the 
idea that $\eta_g$ could be as large as 2,
although because current experiments only have limited/smeared coverage in $x$, 
it is hard to make more exact statements.
In my opinion a definitive measurement of the {\it polarized} gluon will 
only come using a process which has been successfully 
used to determine the {\it unpolarized} gluon. In all likelihood this will require
measurement of direct photon production in polarized scattering at RHIC. 

\section{B-physics}
\begin{figure}
\begin{minipage}{3in}
\begin{center}
\includegraphics[scale=0.42]{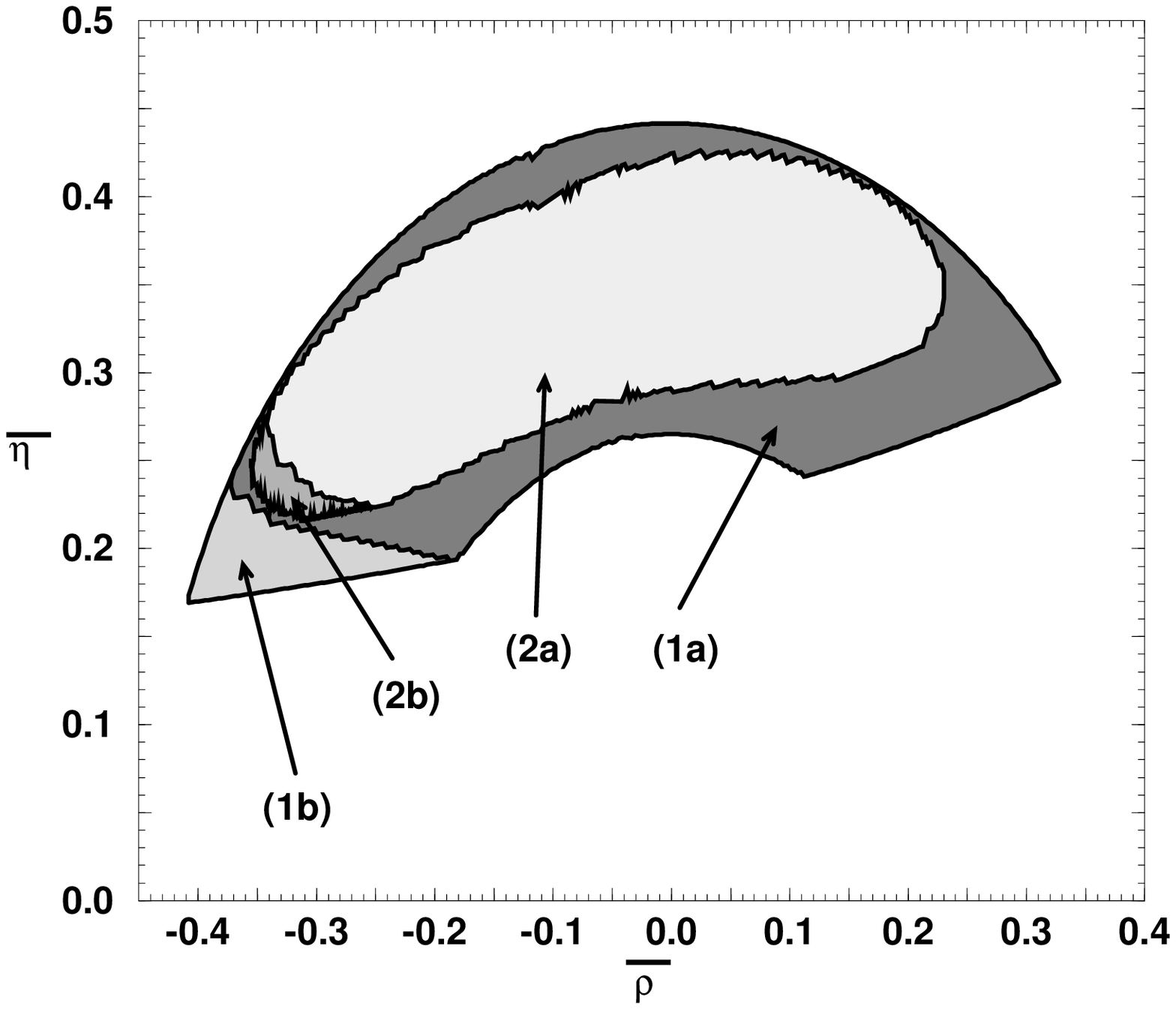}
\end{center}
\end{minipage}
\begin{minipage}{3in}
\begin{center}
\includegraphics[scale=0.40]{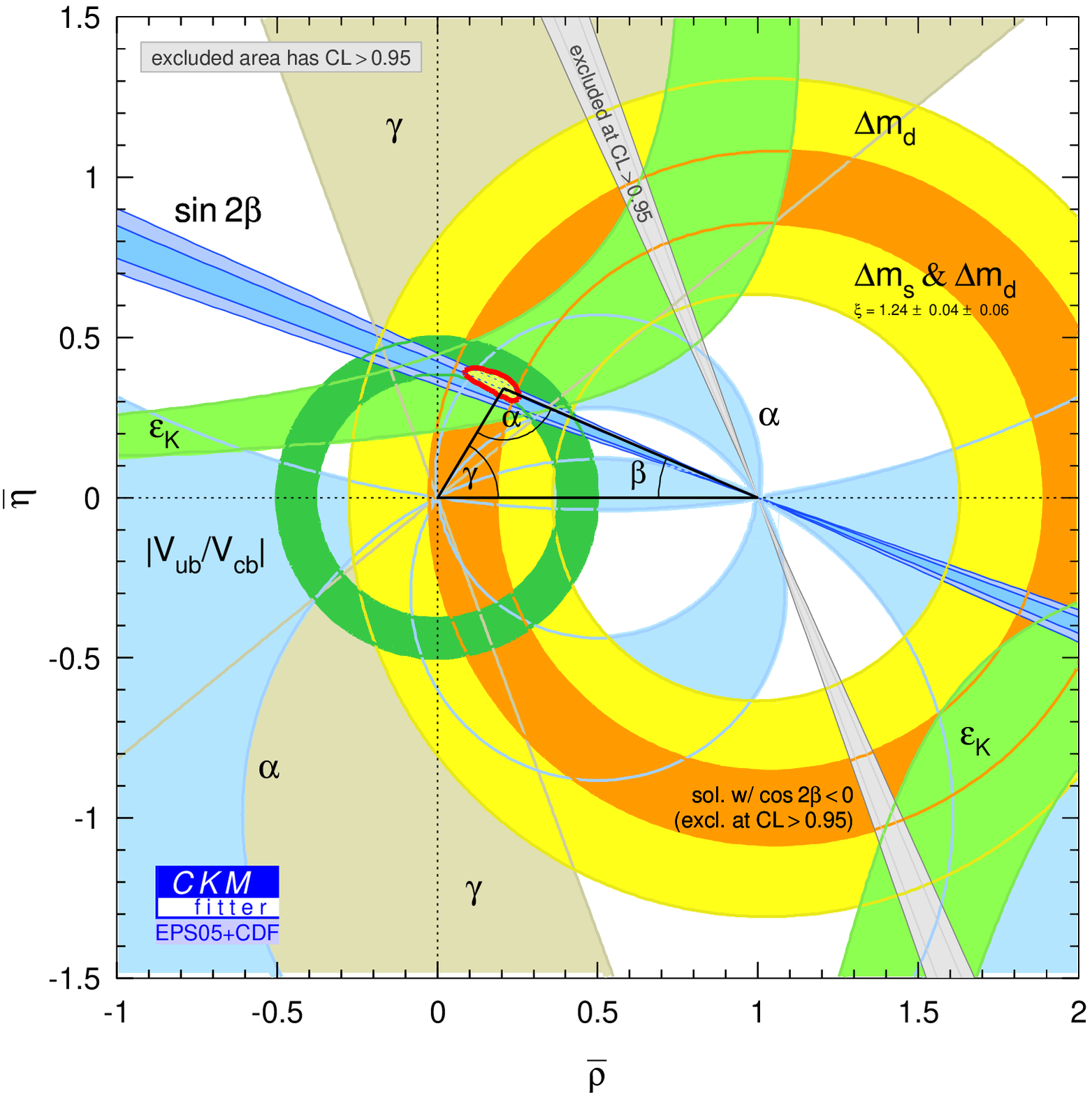}
\end{center}
\end{minipage}
\caption{The constraints on the unitarity triangle in 1995 (left) and in Winter 2006, [16,17].}
\label{Utriangle}
\end{figure}

Fig.~\ref{Utriangle} shows the remarkable progress in the knowledge
of the Unitarity triangle
since the advent of the $B$-factories.
The 1995 plot~\cite{Herrlich:1995hh} contains only the three quantities, $V_{ub},\epsilon_K$ 
and $\Delta m_{B_d}$ and region
1a corresponds to a scan over a $1\sigma$ ranges of the input parameters. The 2006 
plot~\cite{CKMfitter} includes the information on $\Delta m_s$ coming from the Tevatron.
 
\subsection{B-physics at the Tevatron}
The potential of hadron machines to do B-physics stems from the 
large cross section, and from the fact that all types of B-mesons and B-baryons are produced.
At the Tevatron, usable $B$'s are produced at a rate of about 50 Hz.
New results on $B_s$ mixing indicate that the Tevatron is beginning to live up to that potential.
From the D0 study of $B_s$-mixing~\cite{D0,Naimuddin:2006sw}, 
the log likelihood curve has a preferred value at the 
oscillation frequency $\Delta m_s=19~$ps$^{-1}$
with a 90\% confidence level interval $17 < \Delta m_s < 21~$ps$^{-1}$.
D0 are therefore close to an interesting $B_s$ mixing measurement.
After the conference was over, 
CDF~\cite{CDF}  announced a measurement of $\Delta m_{s}$ 
\beq
\Delta m_{s}= 17.31^{+0.33}_{-0.18}(stat) \pm  0.07(sys)~\mbox{ps}^{-1}.
\eeq
On the basis of this measurement CDF deduce that $\frac{|V_{td}|}{|V_{ts}|}  =  
0.208^{+0.001}_{-0.002}\mbox{(stat +sys)}^{+0.008}_{-0.006}\mbox{(theory)}$.

I shall now discuss the source of the theoretical error in this measurement.
Mixing in the neutral $B$ system is determined by box diagrams, leading to the formula
\beq \label{Boxeqn}
\Delta m_{B_q} \sim f_{B_q}^2 \, B_{B_q} \, m_{B_q} \, |V_{tb}^* V_{tq}|^2 
\eeq
Within the CKM model, 
$B_s$ mixing is used primarily to control hadronic uncertainties,
which are present in the decay constant, $f_{B_q}$ and the bag parameter $B_{B_q}$.
We shall see that even with the cancellations inherent in taking the ratio
$\Delta m_{B_d}/\Delta m_{B_s}$ the level
at which the hadronic parameters are currently controlled by lattice gauge theory 
is now inadequate.

\subsection{Lattice results for decay constants, $f_B$ and bag parameters, $B$}
\begin{figure}
\begin{center}
\includegraphics[scale=0.4]{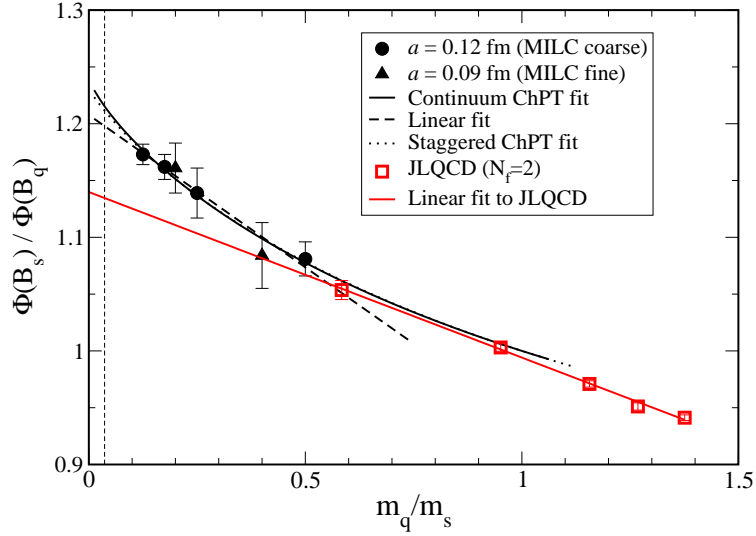}
\caption{Extrapolation of lattice results to the physical mass
using JLQCD and HPQCD data [23].}
\label{phi_sd_all.proc}
\end{center}
\end{figure}

The results from unquenched lattice results QCD are~\cite{Gray:2005ad},
\beq
f_B = 216 (09) (19) (07)~{\rm MeV},\;\; f_{B_s} = 249 (07) (26 )(09)~{\rm MeV}\;.
\eeq
The largest error is due to the fact that matching for the weak current operator 
between the continuum and the lattice is performed at one loop. 
However the matching error (and some other errors) cancel in the ratio, leading to a 4\% prediction.
\beq
\frac{f_{B_s}}{f_{B_d}} = 1.20 (3) (1)\;.
\eeq
The bag parameter is determined by the JLQCD collaboration 
to be~\cite{Aoki:2003xb}
\beq
B_{B_d}(m_b) =0.836 (27) ^{(+56)}_{(-62)},\;\; \frac{B_{B_d}}{B_{B_d}} = 1.017 (16)^{(+56)}_{(-17)}\;.
\eeq
This gives us an 11\% prediction $f_{B_d}\sqrt{B_{B_d}}= 244 (26) {\rm MeV}$,
but a 4\% prediction for the ratio
$\frac{f_{B_s}\sqrt{B_{B_s}}}{f_{B_d}\sqrt{B_{B_d}}}= 1.210 ^{+47}_{-35}$.
In summary the ratio $\xi =\frac{f_{B_d } \sqrt{B_{B_d}}}{f_{B_s } \sqrt{B_{B_s}}}$
is calculated by lattice gauge theory~\cite{Okamoto:2005zg},
\beq
\xi=\frac{\Phi(B_s)}{\Phi(B_d)}
 = 1.210 ^{+0.047}_{-0.035}.
\eeq
Fig.~\ref{phi_sd_all.proc}, taken from ref.~[23],
shows the importance of the low mass points in determining the chiral extrapolation
to the physical mass, shown by the vertical dashed line.
Using this value for the ratio we can extract $|V_{td}/V_{ts}|$ using eq.~(\ref{Boxeqn}),
\beq
\Big|\frac{V_{td}}{V_{ts}}\Big| = \lambda \sqrt{(1-\bar{\rho})^2+\bar{\eta}^2}=
\sqrt{\frac{\Delta m_{B_d}}{\Delta m_{B_s}} }  \;
\sqrt{\frac{ m_{B_s}}{ m_{B_d}} } \; \frac{f_{B_s } \sqrt{B_{B_s}}}{f_{B_d } \sqrt{B_{B_d}}} \; .
\eeq
The 4\% error on $\xi$ has now become the limiting uncertainty, about four times larger
than the error on $\sqrt{\frac{\Delta m_{B_d}}{\Delta m_{B_s}} }$.

\section{QCD engineering and the challenge of the LHC}

\begin{figure}
\begin{center}
\epsfig{file=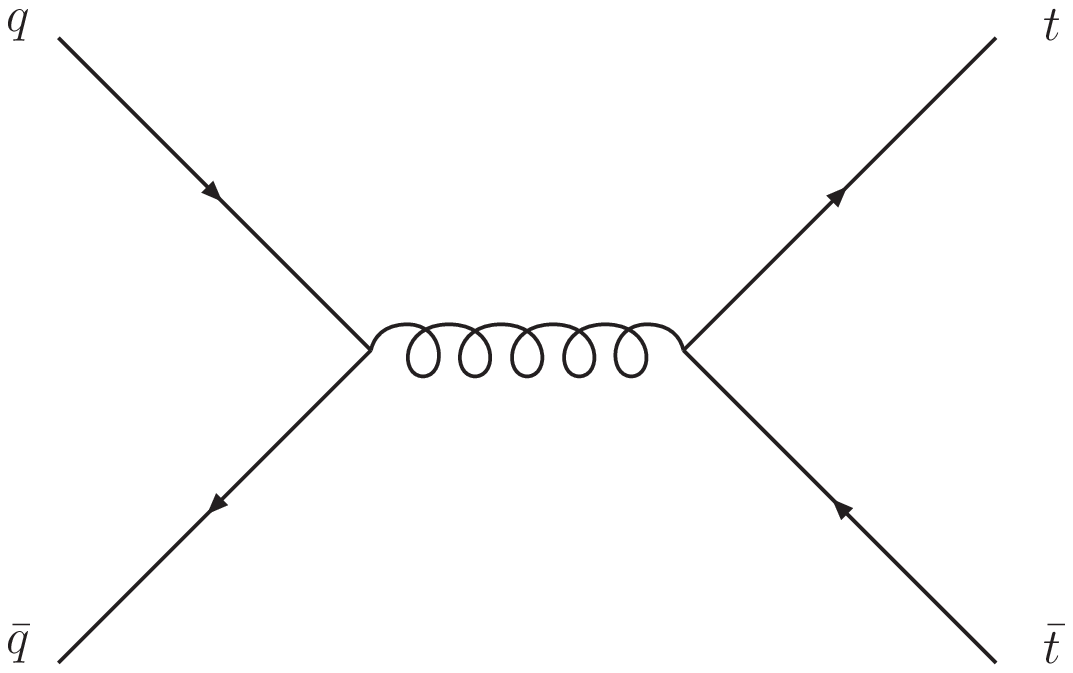,width=4cm,angle=0} \hspace*{2.7cm}
\epsfig{file=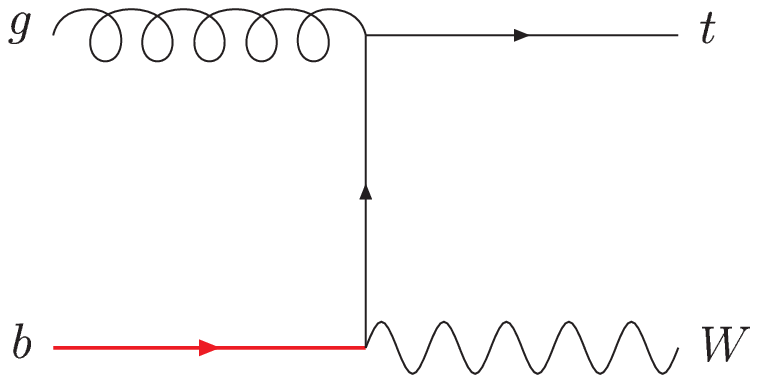,width=4cm,angle=0} \\ \vspace*{0.8cm}
\epsfig{file=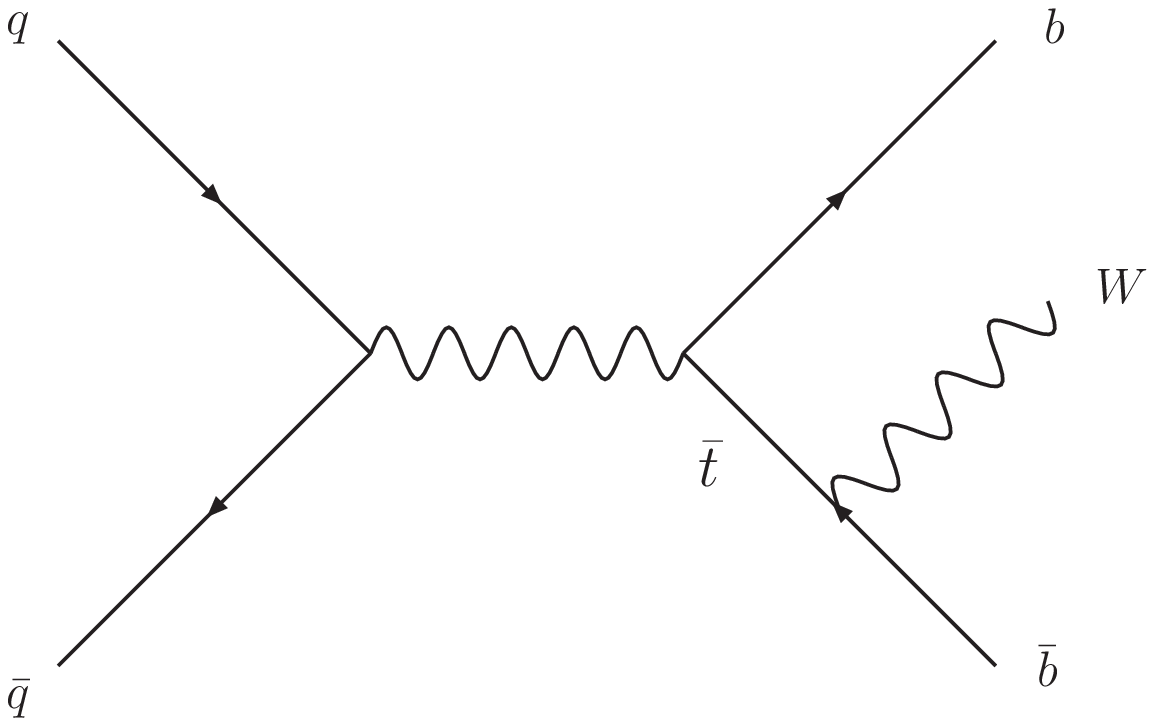,width=4cm,angle=0} \hspace*{2.7cm}
\epsfig{file=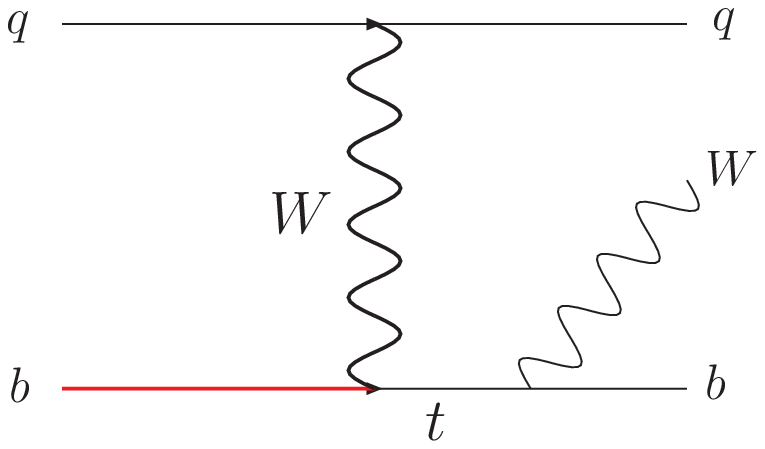,width=4cm,angle=0}
\caption{Lowest order diagrams for pair- and single-top production}
\label{toproddiags}
\end{center}
\end{figure}

\begin{figure}
\begin{center}
\epsfig{file=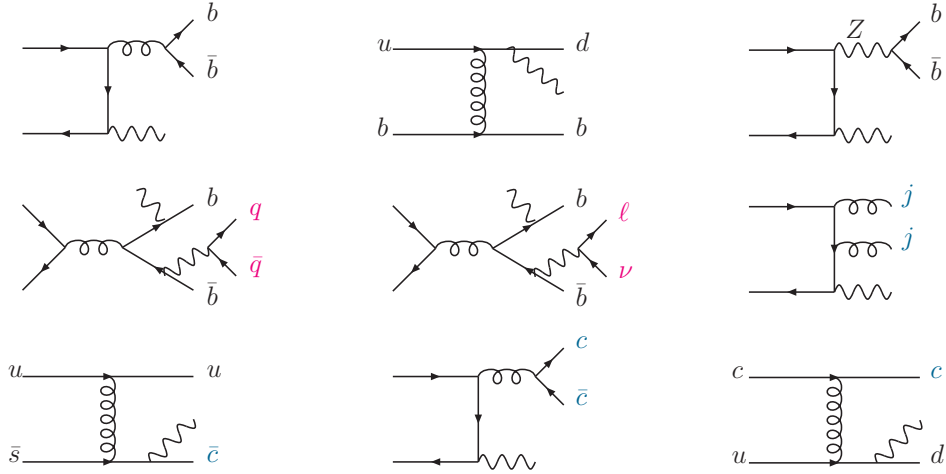,width=11cm,angle=0}
\caption{Backgrounds for single top production}
\label{stbkg}
\end{center}
\end{figure}

In order to assess the challenge presented by the LHC data, 
I shall consider the physics of top production. Since 
some of the motivation for the LHC is to discover objects with large masses,
top production is an interesting paradigm for future studies. Examples of
lowest order diagrams for the pair production of top and for single top 
production are shown in Fig.~\ref{toproddiags}.  
\begin{table}
\begin{center}
\begin{tabular}{|l|c|c|}
\hline
Process & Tevatron[pb] & LHC[pb] \\
\hline
$t \bar{t}$  &  6    & 720 \\
$t \bar{b}$  & 0.8   & 10  \\
$tq$         &  1.8  & 240 \\
$Wt$         & 0.14   & 66 \\
\hline
\end{tabular}
\caption{Top production cross sections at $\sqrt{s}=1.96$~TeV.}
\label{TopXsecs}
\end{center}
\end{table}
Table~\ref{TopXsecs} gives the total next-to-leading order (NLO) 
cross-sections for $t\bar{t}$ production 
both at the Tevatron and the LHC.~\cite{Campbell:2004ch,Campbell:2005bb}
The total single top cross-section is smaller than the $t{\bar t}$ 
rate by about a factor of two, at both machines. However, 
despite the sizeable cross section, 
single top production has not yet been observed at the Tevatron.
This is undoubtedly due to the large number of backgrounds shown in 
Fig.~\ref{stbkg}.
\begin{table}[t]
\begin{center}
\begin{tabular}{|l|c|}
\hline
Process& $\sigma$~[fb]  \\
\hline
$q\bar{q} \to W + b + \bar{b}$ &  30 \\
$q \bar{q} \to W + g + g$ &  35 \\
$u s \to W + u + c$ &  19 \\
$u b \to W + d + b$ &  11 \\
$q\bar{q} \to W + c + \bar{c}$ &  6 \\
$u c \to W + d +c$ &  3 \\
$q \bar{q} \to W + Z(b\bar{b})$ &  3 \\
$q \bar{q} \to t\bar{t}\to Wb\bar{b}q \bar{q}$ &  6 \\
$q \bar{q} \to t\bar{t}\to Wb\bar{b} l\nu$ &  3 \\
\hline
\end{tabular}
\caption{Cross-sections in fb include nominal tagging efficiences and
mis-tagging/fake rates}
\label{bground}
\end{center}
\end{table}
The background processes shown in Fig.~\ref{stbkg}
are calculated with MCFM at $\sqrt{s}=1.96$~TeV and 
the results are given given in Table~\ref{bground}. 
For the cuts and efficiencies used 
we refer the reader to Tables VI,VIII of ref.~[25].
With the same set of cuts, the signal rates are $7$~fb and $11$~fb for 
$s$- and $t-$channel respectively. Thus, with our nominal efficiencies, 
the ratio of signal:background is only $1:6$. 

\begin{figure}
\begin{center}
\includegraphics[angle=270,scale=0.42]{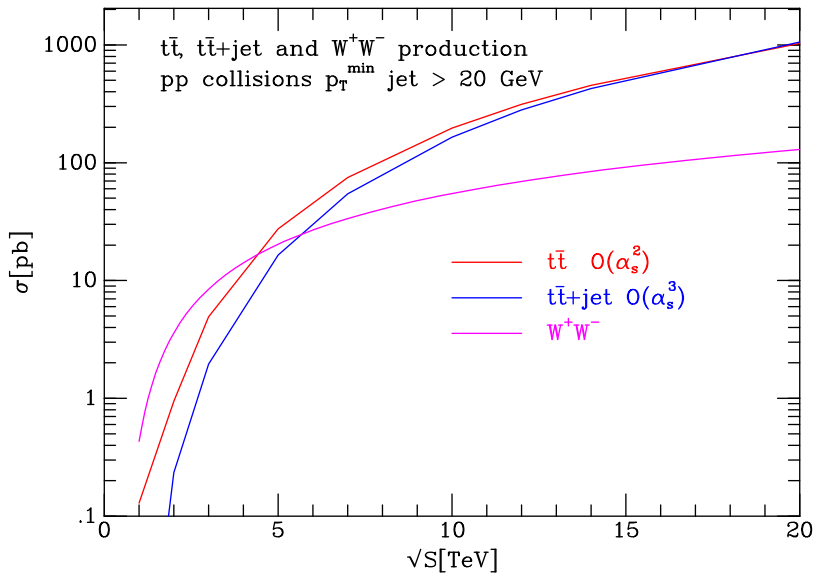}
\caption{Cross section for $t \bar{t}$, $t \bar{t}+\mbox{jet}$ and $W^+W^-$
production as a function of energy}  
\label{toprad}
\end{center}
\end{figure}
Another feature of top production which merits consideration is the production
of jets in association with top quarks. Fig.~\ref{toprad} shows that 
for a jet with $p_T^{min}> 20$~GeV, 
$\sigma(t \bar{t}+{\rm jet}) > \sigma(t \bar{t})$
at the energy of the LHC. Radiation of one gluon in general will not be enough
and the parton shower needs to be included.

These examples make it clear that there is a large amount of work needed to
understand LHC data. Just as in the single top case, signal processes 
will be accompanied by many backgrounds. In the best of all worlds these 
backgrounds will be measured directly in the experiments, 
but in the case of irreducible backgrounds, one will have to rely on 
theoretical calculations. 
 
\begin{figure}
\begin{center}
\includegraphics[angle=270,scale=0.35]{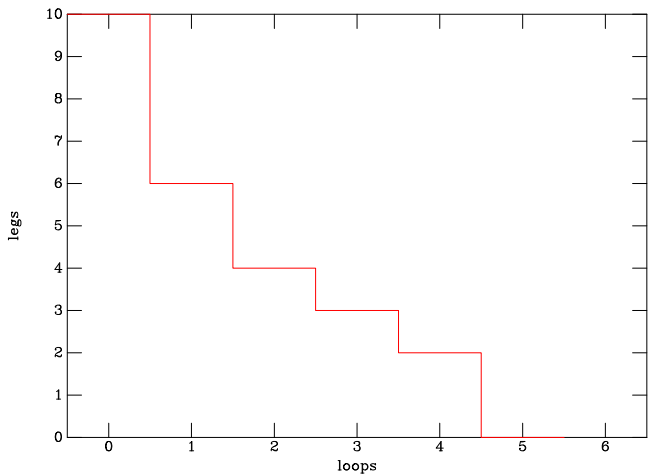}
\caption{Schematic description of the state of the art in multileg calculations}
\label{LL2006}
\end{center}
\end{figure}

The approaches to calculating Feynman amplitudes have been nicely 
reviewed by Del Duca.~\cite{DelDuca:2006cj} The various techniques 
which have been used are tree graphs, tree graphs combined with showers,
NLO calculations, NLO calculations combined with showers, and NNLO calculations.
Specializing for the moment to NLO, Fig.~\ref{LL2006} illustrates the point
that the state of the art can calculate loops or legs, but not both.
 
At LHC processes with many legs will be produced.
Phenemonologically interesting processes involve 
vector bosons, leptons, missing energy, heavy flavours.
Just as in the top case, many processes can contribute 
to the same signature, which argues for a unified approach
to NLO calculations.

The computer program MCFM~\cite{JCampbell}, 
which is a general purpose NLO code,
is an attempt to provide such a tool.
\begin{table}[ht]
\begin{center}
\renewcommand\arraystretch{1.3}
\begin{tabular}{|ll|}
\hline
$p {\bar p} \to W^\pm/Z$ & \qquad  $p {\bar p} \to W^+ + W^-$ \\
$p {\bar p} \to W^\pm + Z$ & \qquad $p {\bar p} \to Z + Z$\\
$p {\bar p} \to W^\pm + \gamma$ &
 \qquad $p {\bar p} \to W^\pm/Z + H$ \\
 $p {\bar p} \to W^\pm + g^\star \, ( \to b {\bar b} )$ &
 \qquad $p {\bar p} \to Zb {\bar b}$ \\
$p {\bar p} \to W^\pm/Z + \mbox{1 jet}$ &
 \qquad $p {\bar p} \to W^\pm/Z + \mbox{2 jets}$ \\
$p {\bar p}(gg) \to H $ & \qquad $p {\bar p}(gg) \to H + \mbox{1 jet}$ \\
$p {\bar p}(VV) \to H + \mbox{2 jets}$ &
 \qquad $p {\bar p} \to t+q$ \\
$p {\bar p} \to H + b$ & \qquad $p {\bar p} \to Z + b$\\
\hline
\end{tabular}
\caption{Processes available in MCFM}
\end{center}
\end{table}
Knowledge of these processes at NLO provides the first precise
predictions of their event rates.
MCFM and other similar programs are a start, 
but they are clearly insufficient for the needs at the LHC.
The stumbling block which prevents the inclusion of further processes is
the calculation of one-loop corrections. 

\section{Techniques for one loop diagrams}
One of the early general calculations of one loop corrections 
was performed by Passarino and Veltman.~\cite{Passarino:1978jh}
More recent general purpose attempts have used semi-numerical results,
which reduce an arbitrary diagram using numerical methods to 
a sum of scalar integrals, which are known analytically. 
Recent results in this field are the one-loop matrix element 
for a  Higgs plus four partons
and the six-gluon amplitudes.~\cite{Zanderighi:2006ud}

\subsection{Analytical techniques}
In addition to the semi-numerical techniques
great progress has been made in the analytical
calculation of loop amplitudes. In this field 
the key ideas are supersymmetric decomposition,
MHV diagrams, BCFW recursion, unitarity
and unitarity using multiple cuts.
Although this is a field which is developing rapidly,
recent reviews are given in refs.~[30,31].

\subsection{Outlook: analytic vs numerical}
The new analytical methods lead to beautiful results 
for gauge theory tree graph amplitudes.
However the evaluation of tree graphs is already solved 
numerically by Berends-Giele recursion.~\cite{Berends:1987me} 
For tree graphs
the issue reduces merely to a question of numerical expediency
which has been addressed in ref.~[33].
So far the impact on real phenomenology is rather limited, although simple tree graph 
results have been obtained for Higgs+5~parton 
amplitudes.~\cite{Dixon:2004za,Badger:2004ty}

The extension of these techniques to loops in QCD is the next frontier.
The new techniques solve the problem of computing one-loop amplitudes of gluons
in ${\cal N}=4$ super Yang-Mills. 
So far only partial analytic results for $n \geq 6$ 
QCD amplitudes have been published.
However there is great intellectual excitement and an injection of personnel 
from formal areas.
There is no doubt that analytic results, when available, are superior 
to numerical, or seminumerical methods. 
It is primarily a question of expediency. 
Which technique will lead first to useful NLO results for the many 
multi-leg processes which we are interested in at the LHC?

\section{Beyond NLO?}
\subsection{NLO and parton showers}
A remarkable theoretical advance of the last few years has 
been the consistent combination of NLO calculations
with parton shower Monte Carlo programs.
The output is a set of events, which are fully inclusive.
Total rates are accurate to NLO in the sense that
NLO results are recovered  for all observables 
upon expansion in $\alpha_S$.
Currently a limited number of available processes, 
single vector boson production, $W/Z/H$,
vector boson pair production, $WW$,
heavy quark pair, $Q\bar{Q}$~\cite{Frixione:2003ei},
and single top production.~\cite{Frixione:2005vw}
It is clear that MC@NLO relies on the appropriate 
NLO process having been calculated.

\begin{figure}
\begin{center}
\includegraphics[scale=0.5]{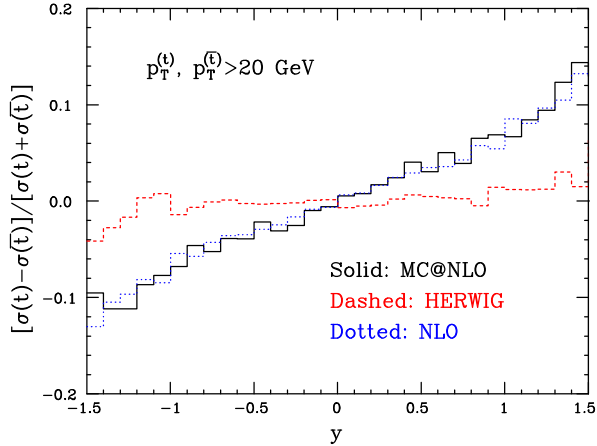}
\end{center}
\caption{Asymmetry in top production at the Tevatron, [36].}
\label{topasy}
\end{figure}

As an example I show a curve obtained for $t \bar{t}$-production 
using MC@NLO.
Figure~\ref{topasy} 
shows the prediction of forward-backward 
asymmetry at the Tevatron.~\cite{Frixione:2003ei}
The asymmetry is generated beyond the 
leading order in perturbation theory~\cite{Nason:1987xz}
and is well reproduced by MC@NLO.

\subsection{What quantities do we want to calculate at NNLO?}
The quantities we need to calculate are those quantities which 
are will be measured at high precision. It is especially important 
to calculate fully differential results, to correctly account for 
experimental cuts. A partial list includes,
\bit
\item $e^+ e^- \to 3$~jets, ($\as$)
\item
Fully exclusive $pp \to W,Z, \gamma^*$, (parton luminosity)
\item
$pp \to 2$~jets (gluon distribution function)
\item 
$pp \to \gamma$+jet (jet energy scale).
\eit
Petriello presented a new result on $W$ production at NNLO including the 
spin correlations.~\cite{Melnikov:2006di}  
For the electron $p_T > M_W/2$ one has effectively a NLO process; corrections
of order 20\% are found. In other regions, 
a perturbation theory uncertainty at the 
per cent level is found.

%%%%%%%%%End MHV
\section{Conclusions}
This week has given testimony to the experimental 
and theoretical vibrancy of QCD. 
Building on persuasive description
of Deep Inelastic data from HERA in the QCD improved dipole model 
theorists have been exploring the structure of the nucleon and the nucleus 
at high energy beyond the saturation limit.
The start of the era of 
precision $B$ physics at hadronic colliders, and, in particular, the
measurement of $B_s$ mixing, presents new challenges for the 
calculation of hadronic parameters from lattice gauge theories.

Lastly, the interpretation of data from LHC will require
puts new demands on QCD calculations, at leading order, 
next-to-leading order (NLO), NLO+parton shower and even at 
next-to-next-to-leading order. It will also require an extensive 
program of validation of these calculations against data and
against themselves.
The LHC begins in about a year, so for the experimenters, 
the era of giving talks about 'Studies at the LHC'
without the benefit of data is almost over. 
The exploration of a completely new range of energy
is a once-in-a-lifetime opportunity for most of us.
Looking further ahead, the continued exploration of the energy
frontier beyond the LHC, requires the success of the LHC experiments.

\section*{Acknowledgments}
I would like to thank Jean Tran Thanh Van and all the Moriond 
staff for their magnificent hospitality in La Thuile.

\end{document}